\documentclass{ceurart-arxiv}

\usepackage{listings}
\usepackage{algorithm}
\usepackage{algorithmicx}
\usepackage[noend]{algpseudocode}
\usepackage{amsmath,mathtools,amsthm}
\usepackage{booktabs}
\usepackage{tikz}
\usepackage{pgfplots}
\usetikzlibrary{patterns}
\usetikzlibrary{arrows.meta}
\usetikzlibrary{arrows.meta,positioning,fit,calc,backgrounds}
\pgfdeclarelayer{annot}
\pgfsetlayers{background,main,annot}
\tikzset{
  lab/.style  = {inner sep=1pt, font=\bfseries\small},
  rec/.style  = {inner sep=1pt, align=left, font=\small},
  nbox/.style = {draw, rounded corners=2pt, fill=white, inner sep=4pt},
  ebox/.style = {fill=white, inner sep=2pt},   
  gedge/.style= {-{Stealth[length=2mm]}, thick},
  ghost/.style= {-{Stealth[length=2.4mm]}, draw=black!35, line width=1pt,
                 line cap=round, shorten >=1pt},
}

\newtheorem{example}{Example}

\newcommand{\recl}[1]{$\begin{aligned}[t]#1\end{aligned}$}
\newcommand{\gnode}[4]{%
  \node[lab, anchor=south] (#1lab) at ($(#2)+(0,0.7mm)$) {#3};
  \node[rec, anchor=north] (#1rec) at ($(#2)-(0,0.7mm)$) {#4};
  \begin{pgfonlayer}{background}
    \node[nbox, fit=(#1lab)(#1rec)] (#1) {};
  \end{pgfonlayer}
}
 
\newcommand{\gedgelab}[5][south]{%
  \node[rec, anchor=#1] (#2rec) at (#3) {#5};
  \node[lab, anchor=south] (#2lab) at (#2rec.north) {#4};
  \begin{pgfonlayer}{background}
    \node[ebox, fit=(#2lab)(#2rec)] (#2) {};
  \end{pgfonlayer}
}

\newcommand{\iris}{\mathsf{IRIs}}
\newcommand{\blanks}{\mathsf{Blanks}}
\newcommand{\lits}{\mathsf{Literals}}
\newcommand{\terms}{\mathsf{Terms}}
\newcommand{\tterms}{\mathsf{Triples}}
\newcommand{\graph}{G}

\newcommand{\nodes}{\mathsf{Nodes}}
\newcommand{\edges}{\mathsf{Edges}}
\newcommand{\keys}{\mathsf{Keys}}
\newcommand{\vals}{\mathsf{Values}}
\newcommand{\labs}{\mathsf{Labels}}
\newcommand{\recs}{\mathsf{Records}}

\newcommand{\str}{\mathsf{Strings}}

\newcommand{\edgf}{\mathsf{edg}}
\newcommand{\labf}{\mathsf{lab}}
\newcommand{\recf}{\mathsf{rec}}
\newcommand{\ntfun}{\textsc{id}}

\newcommand{\rubenvar}{\textsc{Structural}}
\newcommand{\maximevar}{\textsc{Direct}}
\newcommand{\katjavar}{\textsc{Fold}}

\newcommand{\ireifies}{\mathit{reifies}}
\newcommand{\isun}{\mathit{sun}}
\newcommand{\imadeof}{\mathit{madeOf}}
\newcommand{\istatedby}{\mathit{statedBy}}
\newcommand{\ialice}{\mathit{alice}}
\newcommand{\ibob}{\mathit{bob}}
\newcommand{\iplasma}{\mathit{plasma}}
\newcommand{\ipopcorn}{\mathit{popcorn}}
\newcommand{\icertainty}{\mathit{certainty}}

\newcommand{\kasserted}{\mathsf{asserted}}
\newcommand{\krefids}{\mathsf{refId}}
\newcommand{\krefid}{\krefids}
\newcommand{\kin}{\mathsf{in}}
\newcommand{\kreifiedby}{\mathsf{reifiedBy}}
\newcommand{\kreferences}{\mathsf{references}}
\newcommand{\krefprop}{\mathsf{refProperty}}

\begin{document}

\copyrightyear{2026}
\copyrightclause{Copyright for this paper by its authors.
  Use permitted under Creative Commons License Attribution 4.0
  International (CC BY 4.0).}

\conference{DMKG’26: 2nd International Workshop on Data Management for Knowledge Graphs, October 2026, Bari, Italy}

\title{Translating RDF 1.2 Graphs to Property Graphs}

\author[1]{Maxime Jakubowski}[%
orcid=0000-0002-7420-1337,
email=maxime.jakubowski@tuwien.ac.at,
]
\address[1]{TU Wien, Austria}
\cormark[1]

\author[2]{Dominik Tomaszuk}[%
orcid=0000-0003-1806-067X,
]
\address[2]{University of Bialystok, Poland}

\author[3]{Daniel Fernández-Álvarez}[%
orcid=0000-0002-8666-7660,
email=fernandezalvdaniel@uniovi.es
]
\address[3]{University of Oviedo, Spain}

\author[4]{Ruben Taelman}[%
orcid=0000-0001-5118-256X,
]
\address[4]{Department of Electronics and Information Systems, Ghent University – imec, Belgium}

\author[4]{Ruben Dedecker}[%
orcid=0000-0002-3257-3394,
]

\author[3]{Jose-Emilio Labra-Gayo}[%
orcid=0000-0001-8907-5348,
]

\author[1]{Katja Hose}[%
orcid=0000-0001-7025-8099,
]
 
\cortext[1]{Corresponding author.}

\begin{abstract}
  RDF 1.2 introduces statement-level constructs, such as triple terms, reifiers, and descriptions, that pose new challenges for transforming RDF data into labeled property graphs (LPGs). This work presents and compares three algorithms for translating RDF 1.2 data into LPGs, each interpreting a different amount of the RDF 1.2 reification vocabulary.
  Algorithm \rubenvar\ interprets no vocabulary and losslessly supports arbitrary RDF 1.2 graphs by representing triple terms explicitly as nodes; \maximevar\ interprets \texttt{rdf:reifies} to recover annotated edges while preserving reifiers as graph objects; and \katjavar\ additionally folds reifier descriptions into edge properties, resulting in a more concise, albeit lossy, LPG representation.
  We define the three translations, analyze which class of RDF 1.2 graphs each preserves, and experimentally evaluate their implementations. The experiments confirm the predicted trade-offs: all variants scale linearly on the tested workloads, the generality of \rubenvar\ has no runtime penalty, and folding descriptions into edge records can significantly reduce the number of edges on description-heavy data.
\end{abstract}


\maketitle

\section{Introduction}
\label{sec:intro}
Graph-structured data has become an increasingly important means of representing and integrating complex information across a wide range of application domains. As graph-based applications continue to grow in scale and complexity, so does the need to exchange data between different graph data management systems and data models.

RDF and labeled property graphs (LPGs) are the two predominant models for graph-structured data. Although both represent information as nodes connected by edges, they differ in how they model relationships and their associated metadata. In many applications, edges/relationships carry additional metadata, such as provenance, confidence scores, timestamps, or sources. LPGs support this natively, since edges may themselves have properties. RDF, by contrast, has no native mechanism for attaching metadata directly to a triple, motivating a long line of work on statement-level metadata, most prominently RDF-star~\cite{hartig2017foundations}.

RDF~1.2, now at Candidate Recommendation status~\cite{rdf12concepts}, standardizes a new mechanism for capturing such meta. Its central construct is the \emph{triple term}, an RDF triple used as a term in the object position of another triple. Metadata is attached through a \emph{reifier}, a resource linked to a triple term by the dedicated predicate \texttt{rdf:reifies}. Interoperability with property graphs was an explicit motivation for the W3C RDF \& SPARQL Working Group chartered to standardize RDF-star~\cite{rdfstarwgcharter}. 
At first sight, the correspondence appears straightforward: the edge-with-properties construct that LPGs provide natively has a natural counterpart in RDF through the combined use of reifiers and triple terms.

The correspondence is nonetheless less direct than it first appears, as RDF~1.2 permits configurations that an LPG cannot represent. A triple term may be the object of any predicate, not only \texttt{rdf:reifies}; it may contain another triple term, nesting to arbitrary depth; and the triple it represents need not itself be asserted, that is, included as a statement in the RDF graph. An LPG edge is more rigid: it cannot be the endpoint of another edge, and it cannot be nested. The apparent correspondence between an annotated triple and an edge with properties is therefore a modeling pattern that RDF~1.2 data may follow, not one guaranteed by the data model. Translating RDF~1.2 into LPGs consequently requires deciding how to handle graphs that depart from this pattern.

Figure~\ref{lst:intro} illustrates three challenges that arise from this mismatch.

\textbf{Referenceability.} The triple \texttt{:r1 :statedBy :alice} states that \texttt{:sun :madeOf :popcorn} was stated by Alice. Ideally, this annotation should be stored in an edge's record. However, record entries are plain values, whereas \texttt{:alice} denotes a graph object in its own right (and might be an endpoint of other edges). Descriptions thus relate statements to graph objects, while edge properties hold values; a translation must bridge this gap, for instance by references from records to nodes.

\textbf{Assertedness.} The first statement of Figure~\ref{lst:intro} is written with \texttt{\textasciitilde:r2}, a shorthand that both asserts the triple (i.e., it also includes it as a statement of the graph) and reifies it. 
In contrast, \texttt{rdf:reifies} alone simply refers to a triple term without asserting it: the triple reified by \texttt{:r1} and \texttt{:r3} is not asserted. Whether a triple term is asserted is thus independent of its reification. An LPG edge has no such freedom; it is simply present. So, a translation must resolve the distinction one way or another: record assertedness explicitly, encode it structurally, or discard it.

\textbf{Multiplicity.} Both \texttt{:r1} and \texttt{:r3} reify the same triple term. More generally, RDF~1.2 allows several reifiers to relate to the same triple term, and a single reifier to relate to several triple terms. The correspondence between reifiers and triple terms is 
many-to-many. 
Then, interpreting every reifier as a distinct LPG edge is a modeling assumption rather than a consequence of the RDF~1.2 data model.

\begin{figure*}[tbp]
  \small
\begin{verbatim}
:sun :madeOf :plasma ~ :r2 .

:r1 rdf:reifies <<( :sun :madeOf :popcorn )>> ;
    :statedBy :alice ;
    :certainty 0.8 .

:r3 rdf:reifies <<( :sun :madeOf :popcorn )>> ;
    :statedBy :bob .
\end{verbatim}
  \caption{An RDF~1.2 graph in Turtle. The first statement is asserted and
  carries the reifier \texttt{:r2}; the proposition that the sun is made of
  popcorn is \emph{not} asserted, and is reified twice.}
  \label{lst:intro}
\end{figure*}

These challenges expose a central design choice: \emph{how much of the RDF~1.2 reification vocabulary should a translation interpret?}
At one extreme, a translation interprets none of it and represents every triple term as a node: the complete graph is preserved, but the LPG serves as a generic container in which no annotation is ever stored as an edge property. At the other extreme, a translation interprets \texttt{rdf:reifies} and folds annotations into edge records, where an LPG natively keeps them, but, at the price of restricting the inputs and discarding information that does not fit.

We investigate this design space through three translations. \rubenvar\ interprets no vocabulary and losslessly supports arbitrary RDF~1.2 graphs by representing triple terms explicitly. \maximevar\ interprets \texttt{rdf:reifies}, representing each reified triple as an edge while preserving reifiers as graph objects. \katjavar\ additionally folds reifier descriptions into edge records (which is the native LPG representation of annotated statements) at the cost of reifier identity. We formally define the translations, analyze their trade-offs, and evaluate them experimentally.

Translation between RDF and property graphs has been studied extensively in both directions. Angles et al.~\cite{angles2020mapping} define three direct RDF-to-property-graph mappings, two of which preserve both information and semantics, while Rabbani et al.~\cite{rabbani2024transforming} additionally exploit SHACL and PG-Schema during translation.
Hartig~\cite{hartig2014reconciliation} formalizes the property graph model and defines transformations in both directions; for the opposite direction specifically, Das et al.~\cite{das2014tale} represent property graphs as RDF in Oracle, including a reification-based encoding.
Closest to our setting, Abuoda et al.~\cite{abuoda2022transforming} analyze transformations from RDF-star to property graphs and identify the cases in which existing approaches fail. Our work differs in targeting RDF~1.2. Previous approaches consider RDF~1.1 or RDF-star, where quoted triples correspond more directly to annotated LPG edges. RDF~1.2 replaces that design with reifiers, restricts triple terms to object position, and gives \texttt{rdf:reifies} a generic interpretation.

The remainder of this paper is organized as follows. Section~\ref{sec:prelim} introduces the required notation. Section~\ref{sec:translations} presents the 
translation approaches, each illustrated using the running example of Figure~\ref{lst:intro}. Section~\ref{sec:evaluation} evaluates the approaches, and Section~\ref{sec:conclusion} concludes this paper with an outlook to future work.


\section{Preliminaries}
\label{sec:prelim}
We briefly recall the two data models involved: RDF~1.2 graphs 
and labeled property graphs.

\paragraph{RDF 1.2 Graphs}
Let $\iris$, $\blanks$, and $\lits$ be mutually disjoint, countably infinite sets of \emph{IRIs}, \emph{blank nodes}, and \emph{literals}; these are the \emph{basic RDF terms}. To account for RDF~1.2's nested triples, the set of \emph{RDF terms} $\terms$ and the set of \emph{RDF triples} $\tterms$ are defined by mutual recursion. An RDF triple $(s, p, o)$ has a \emph{subject} $s \in \iris \cup \blanks$, a \emph{predicate} $p \in \iris$, and an \emph{object} $o \in \terms$; that is, $\tterms \subseteq (\iris \cup \blanks) \times \iris \times \terms$. An RDF term is an IRI, a blank node, a literal, or a triple: $\terms = \iris \cup \blanks \cup \lits \cup \tterms$. An \emph{RDF graph} $\graph$ is a finite set of RDF triples, $\graph \subseteq \tterms$. A triple that is an element of $\graph$ is said to be \emph{asserted} in $\graph$, sometimes also called a \emph{top-level} triple.
 
\paragraph{Triple Terms and Reification}
An RDF triple that occurs as the object of another triple is called a \emph{triple term}. The same triple may occur as an asserted triple, as a (nested) triple term, or both; the two are independent. Indeed, a nested triple need not be asserted. RDF~1.2 uses the IRI $\ireifies \in \iris$\footnote{The full $\ireifies$ IRI is \url{http://www.w3.org/1999/02/22-rdf-syntax-ns\#reifies}} in the predicate of so-called \emph{reifing triples}. 
A \emph{reifying triple} is a triple $(r, \ireifies, \tau)$ whose object $\tau$ is a triple term and whose subject $r$ is called the \emph{reifier}
of $\tau$~\cite{rdf12concepts}. Reifier $r$ is used in further triples that describe $\tau$. We call a triple $(r, p, o)$ with $p \neq \ireifies$, whose subject $r$ is a reifier of $\tau$, a \emph{description triple} of $\tau$.
 
\paragraph{Labeled Property Graphs}  
Let $\nodes$ and $\edges$ be mutually disjoint, countably infinite sets representing \emph{Nodes} and \emph{Edges}, respectively. Let $\keys$ be a countably infinite set of \emph{property keys}, and let $\vals$ be a countably infinite set of \emph{property values}. The set of all \emph{Labels} is denoted by $\labs$, and the set of all \emph{Records} (finite mappings from keys to values) is defined as $\recs = \{ r : K \rightarrow \vals \mid K \subseteq \keys \}$. A \emph{labeled property graph} $\graph$ is defined as a tuple $\graph = (N, E, \edgf, \labf, \recf)$ where:
\begin{itemize}
\item $N \subseteq \nodes$ is a finite set of nodes; $E \subseteq \edges$ is a finite set of edges such that
  $N \cap E = \emptyset$;
\item $\edgf : E \rightarrow (N \times N)$ is a total function mapping
  each edge to an ordered pair of nodes, where for $e \in E$ with
  $\edgf(e) = (u, v)$, we call $u$ the \emph{source} and $v$ the
  \emph{target} of $e$;
\item $\labf : (N \cup E) \rightarrow 2^{\labs}$ is a total function
  mapping nodes and edges to finite sets of labels (including the
  empty set);
\item $\recf : (N \cup E) \rightarrow \recs$ is a total function
  mapping nodes and edges to records (their \emph{properties}).
\end{itemize}
We say a node or edge $o$ \emph{carries} a property key $k$ when $k$ is defined in $\recf(o)$, and that it \emph{carries} $k \mapsto v$ when additionally $\recf(o)(k) = v$.

\section{Translating RDF 1.2 to LPGs}
\label{sec:translations}
The three translations share the same principle: every basic RDF term occurring as a subject or object becomes a node (literals included, so
a literal is represented once as a node shared by all triples mentioning it, rather than inlined as a value) and every triple becomes an edge, labeled by its predicate. Throughout, blank nodes are Skolemized and treated as IRIs.
The translations deviate from this only in their treatment of triple terms, reifiers, descriptions, and assertedness (cf. Table~\ref{tab:tradeoffs}). \rubenvar\ interprets no vocabulary. \maximevar\ and \katjavar\ interpret $\ireifies$, representing each reified triple as an edge: \maximevar\ keeps reifiers as nodes referenced from these edges, while \katjavar\ folds a reifier's descriptions into the edge record itself, the native LPG representation of an annotated statement.

\paragraph{From RDF terms to LPG nodes}
We reconcile the two data models with a single naming convention. Writing $\str$ for the set of all finite strings, we assume $\str \subseteq \nodes \cap \vals \cap \labs$, so that one string may serve as a node, as a property value, and as a label. Fix an injective function $\ntfun : \terms \to \str$ that assigns every RDF term a distinct string identifier\footnote{An example of such a function is the N-Triples serialization.}. Since $\str \subseteq \nodes$, the identifier $\ntfun(t)$ is itself a node, and we take it as the node representing the RDF term $t$. Indeed, injectivity is what makes node identity sound: two occurrences of the same term yield the same node.

Our translations reference nodes from within edge records. However, concrete systems keep node identifiers internal and do not expose them as addressable values~\cite{isogql}. To keep the translations realizable, every node additionally carries its identifier as an ordinary property under the reserved key $\krefid$: the node of $t$ carries $\krefid \mapsto \ntfun(t)$, and edges never carry $\krefid$. A reference is then an equality between property values, in the manner of a \emph{foreign key}: a referring key, carried on an edge, holds the $\krefid$ of the node it denotes. Each variant has its own referring keys: $\kin$ for \rubenvar, $\kreifiedby$ for \maximevar, and, for \katjavar it is $\kreferences$.

Predicates, in contrast, become labels: the predicate $p$ of a triple becomes the label $\ntfun(p) \in \labs$ of its edge (an IRI occurring both as a predicate and as a subject or object thus results in both a label and a node). Finally, two helpers, shared by all algorithms, type and populate the node of a term $t$. $\textsc{Kind} : \terms \to \labs$ returns $\mathsf{IRI}$, $\mathsf{Literal}$, or $\mathsf{TripleTerm}$ according to whether $t \in \iris$, $t \in \lits$, or $t \in \tterms$. $\textsc{Record} : \terms \to \recs$ returns $\{\krefid \mapsto \ntfun(t)\}$, extended for literals with $\mathsf{value}$ and $\mathsf{datatype}$ (plus $\mathsf{language}$ or $\mathsf{direction}$ for language-tagged and directional strings).

\subsection{Translation \rubenvar}
 

\rubenvar\ interprets no RDF~1.2 vocabulary; in particular, it treats $\ireifies$ as an ordinary predicate. Each triple term $\tau$ is materialized as a node $\ntfun(\tau)$ labeled ``$\mathsf{TripleTerm}$''. Being a node, $\ntfun(\tau)$ can be pointed at by edges, so any triple whose object is $\tau$ (e.g., a reifying triple $(r, \ireifies, \tau)$ or otherwise) is just a plain edge into $\ntfun(\tau)$, with no special handling. Then, to represent the triple $\tau = (s,p,o)$ itself, we emit it as an edge $e$ from $s$ to $o$ labeled $p$. However, $e$ must be distinguished from an assertion and it must be tied back to $\ntfun(\tau)$. Since we cannot simply create an edge from $e$ to $\ntfun(\tau)$ itself (that would mean making $e$ an endpoint of another edge, which LPGs forbid), 
\rubenvar\ uses the $\krefid$ introduced above. 
The edge key $\kin \in \keys$ is added to the record of $e$ which carries the reference to $\ntfun(\tau)$
Furthermore, an edge with $\kin \mapsto \ntfun(\tau)$ as part of its record represents the triple term $\tau$, whereas an edge with \emph{no} $\kin$ key represents an asserted triple of $\graph$. Assertedness is thus recorded structurally; one can verify assertedness by checking the presence of $\kin$.

Throughout this section we illustrate each translation on the running example of Figure~\ref{lst:intro}. We build it up gradually: we begin with the small
fragment that already exercises the essential case for \rubenvar, then add the
asserted triple, the second description, and the second reifier as they become
relevant for \maximevar\ and \katjavar.

\begin{figure}
  \footnotesize
\centering
\begin{tikzpicture}[x=1cm, y=1cm]
 
  \gnode{alice}  {-6,0}{IRI}     {$\krefids\mapsto \ntfun(\ialice)$}
  \gnode{r1}     {0,0}{IRI}     {$\krefids\mapsto \ntfun(r_1)$}
  \gnode{sun}    {0,1.5}{IRI}     {$\krefids\mapsto \ntfun(\isun)$}
  \gnode{t1}     {6,0}{TripleTerm}     {$\krefids\mapsto \ntfun(\tau_1)$}
  \gnode{popcorn}{6,1.5}{IRI}     {$\krefids\mapsto \ntfun(\ipopcorn)$}
 
  \draw[gedge] (r1)  -- (alice);
  \draw[gedge] (r1)  -- (t1);
  \draw[gedge] (sun) -- (popcorn);
 
  \gedgelab      {e1}{-3,-0.1}{statedBy}{}
  \gedgelab      {e2}{3,0}{reifies}{}
  \gedgelab      {e3}{3,1.5}{madeOf}{$\kin\mapsto \ntfun(\tau_1)$}
 
  \begin{pgfonlayer}{annot}
    \draw[ghost,dashed] ([xshift=-3mm] e3rec.south east) to[out=-90, in=180] (t1rec.west);
  \end{pgfonlayer}
 
\end{tikzpicture}
\caption{The \rubenvar\ encoding of Example~\ref{ex:ruben}. Solid
  arrows are LPG edges. The gray dashed arrow indicates the foreign key
  reference.}
\label{fig:ruben-example}
\end{figure}
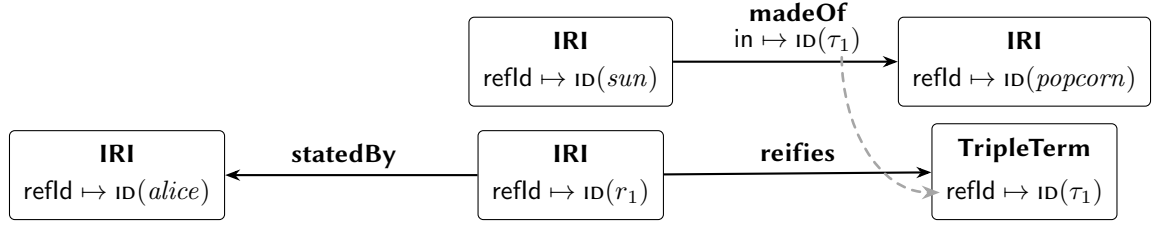

\begin{example}\label{ex:ruben}
  We start from the smallest interesting fragment of Figure~\ref{lst:intro}. Let $\tau_1 = (\isun,\imadeof,\ipopcorn)$ and $\graph_\rubenvar = \{(r_1,\ireifies,\tau_1),(r_1,\istatedby,\ialice)\}.$ Here $r_1$ reifies $\tau_1$,  and the description $(r_1,\istatedby,\ialice)$ records who stated it. Note that $\tau_1$ is \emph{not} asserted: it occurs only as a triple term, never as a top-level triple of $\graph_\rubenvar$. \rubenvar\ creates five nodes: $\ntfun(\isun)$, $\ntfun(r_1)$, $\ntfun(\ialice)$, and $\ntfun(\ipopcorn)$ labeled $\mathsf{IRI}$, and $\ntfun(\tau_1)$ labeled $\mathsf{TripleTerm}$, and three edges: two for the triples of $\graph_\rubenvar$, and one carrying $\kin\mapsto\ntfun(\tau_1)$ for the content of $\tau_1$. Figure~\ref{fig:ruben-example} shows the output $\rubenvar(G_\rubenvar)$.
\end{example}

Algorithm~\ref{alg:ruben} has a shared state consisting of the five LPG components and an additional set $\textit{encoded}$ across all procedure calls. The core procedure \textsc{Emit} turns one triple into one edge; its second argument records the context of that triple, either $\bot$ (an asserted triple, top level) or the identifier of the enclosing triple term. When the object is itself a triple term $\tau$ that has not yet been emitted, \textsc{Emit} recurses on $\tau$ to emit its triple; deeper nesting is handled by the same recursion. The guard on line~12 emits each triple term exactly once, so a triple term with several reifiers keeps its distinct $\ireifies$ edges (one per reifier) while sharing a single node with label `$\mathsf{TripleTerm}$' and a single copy of its triple.
 
\begin{algorithm}[h]
  \small
\caption{\rubenvar: interprets no vocabulary}
\label{alg:ruben}
\begin{algorithmic}[1]
  \Statex \textbf{Shared state:} $N, E, \edgf, \labf, \recf, \textit{encoded}$
  \Procedure{\rubenvar}{$\graph$}
    \State $N \gets \emptyset$;\quad $E \gets \emptyset$;\quad
           $\edgf, \labf, \recf \gets$ empty maps;\quad
           $\textit{encoded} \gets \emptyset$
    \ForAll{$(s, p, o) \in \graph$}
      \State \textsc{Emit}$\big((s, p, o),\ \bot\big)$
        \Comment{$\bot$: this is a graph triple}
    \EndFor
    \State \Return $(N, E, \edgf, \labf, \recf)$
  \EndProcedure
 
  \Statex
  \Procedure{Emit}{$(s, p, o),\ \textit{in}$}
      \Comment{$\textit{in} \in \vals \cup \{\bot\}$}
    \State \textsc{AddNode}$(s)$;\quad \textsc{AddNode}$(o)$
    \State choose $e \in \edges \setminus E$;\quad $E \gets E \cup \{e\}$
    \State $\edgf(e) \gets \big(\ntfun(s),\ \ntfun(o)\big)$
    \State $\labf(e) \gets \{\ntfun(p)\}$
      \Comment{the predicate is the edge label}
    \State $\recf(e) \gets
      \begin{cases}
        \emptyset & \text{if } \textit{in} = \bot\\
        \{\kin \mapsto \textit{in}\} & \text{otherwise}
      \end{cases}$
    \If{$o \in \tterms$ \textbf{and} $\ntfun(o) \notin \textit{encoded}$}
      \Comment{triple term whose triple is not yet emitted}
      \State $\textit{encoded} \gets \textit{encoded} \cup \{\ntfun(o)\}$
      \State \textsc{Emit}$\big(o,\ \ntfun(o)\big)$
        \Comment{emit its triple; nesting handled by $o$'s object}
    \EndIf
  \EndProcedure
 
  \Statex
  \Procedure{AddNode}{$t$}
      \Comment{materialize $t$ as a node}
    \If{$\ntfun(t) \notin N$}
      \State $N \gets N \cup \{\ntfun(t)\}$
      \State $\labf(\ntfun(t)) \gets \{\textsc{Kind}(t)\}$
      \State $\recf(\ntfun(t)) \gets \textsc{Record}(t)$
    \EndIf
  \EndProcedure
\end{algorithmic}
\end{algorithm}

\subsection{Translation \maximevar} \label{sec:maximevar}
\maximevar\ makes two assumptions about its input: (i)~a triple term occurs only as the object of a $\ireifies$ triple; (ii)~triple terms are not nested. We call an RDF~1.2 graph satisfying both a \emph{flat (reification) graph}. The assumptions are imposed by the structure of LPG edges: an edge cannot be the endpoint of another edge, nor can it contain one, so a triple term that is the object of an arbitrary predicate, or that nests another triple term, cannot be represented as an edge at all. In flat graphs, every triple term can be represented as an edge: it is an RDF term using basic RDF terms, occuring only in reifying triples. The excluded nesting structures are exactly those that \rubenvar\ handles by turning triple terms into nodes; \maximevar\ targets flat reification graphs and leaves the rest to \rubenvar.

\maximevar\ interprets $\ireifies$: where \rubenvar\ keeps a reifying triple $(r, \ireifies, \tau)$ as an edge into a $\mathsf{TripleTerm}$ node, \maximevar\ represents the triple term $\tau = (s,p,o)$ itself as an edge from $s$ to $o$ labeled $p$, which we call a \emph{triple term edge}. Such an edge is distinguished from an ordinary one by carrying $\kreifiedby \mapsto \ntfun(r)$, a foreign key to the node of its reifier $r$ (recall that $\ntfun(r)$ carries $\krefid$). The reifier $r$ always becomes a node labeled $\mathsf{Reifier}$, serving as the target of this foreign key.

\begin{figure}
\centering
\begin{tikzpicture}[x=1cm, y=0.8cm]
 
  \gnode{plasma} {0,1.8}{IRI}     {$\krefids\mapsto \ntfun(\iplasma)$}
  \gnode{sun}    {6,1.8}{IRI}     {$\krefids\mapsto \ntfun(\isun)$}
  \gnode{popcorn}{12,1.8}{IRI}    {$\krefids\mapsto \ntfun(\ipopcorn)$}
  \gnode{alice}  {13,0}{IRI}    {$\krefids\mapsto \ntfun(\ialice)$}
  \gnode{r2}     {1,0}{Reifier} {$\krefids\mapsto \ntfun(r_2)$}
  \gnode{r1}     {7,0}{Reifier} {$\krefids\mapsto \ntfun(r_1)$}
  \draw[gedge] (sun) -- (plasma);
  \draw[gedge] (sun) -- (popcorn);
  \draw[gedge] (r1) -- (alice);
 
  \gedgelab      {e1}{3.2,1}{madeOf}{\recl{
      \kasserted &\mapsto \mathit{true}\\
      \kreifiedby &\mapsto \ntfun(r_2)\\
    }}
  \gedgelab      {e2}{8.8,1}{madeOf}{\recl{
      \kasserted &\mapsto \mathit{false}\\
      \kreifiedby &\mapsto \ntfun(r_1)\\
    }}
  \gedgelab      {e3}{9.8,-0.5}{statedBy}{$\kasserted \mapsto \mathit{true}$}
 
  \begin{pgfonlayer}{annot}
    \draw[ghost,dashed] ([xshift=-5.5mm] e1rec.south east) to[out=-90, in=0] (r2rec.east);
    \draw[ghost,dashed] ([xshift=-6mm] e2rec.south east) to[out=-90, in=0] (r1rec.east);
  \end{pgfonlayer}
\end{tikzpicture}
\vspace{-2em}
\caption{The \maximevar\ encoding of Example~\ref{ex:maxime}.}
\label{fig:maxime-example}
\end{figure}
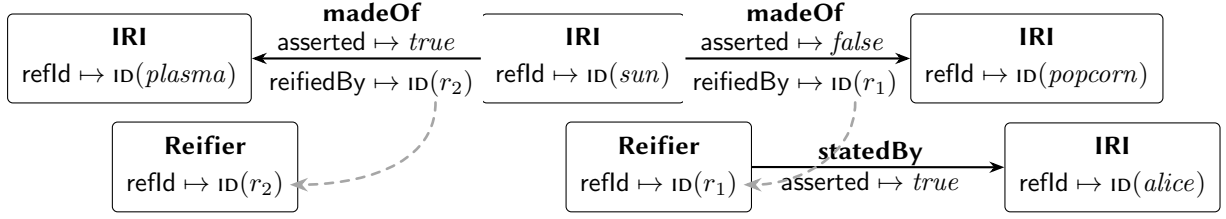

\begin{example}\label{ex:maxime}
  We extend $G_\rubenvar$ of Example~\ref{ex:ruben} with an asserted triple and its reifier. Let $\tau_2 = (\isun,\imadeof,\iplasma)$ and $G_\maximevar = G_\rubenvar \cup \{(\isun,\imadeof,\iplasma), (r_2,\ireifies,\tau_2)\}$.
  In contrast to $\tau_1$, the triple $\tau_2$ is \emph{asserted}: it occurs both as the triple term reified by $r_2$ and as a plain triple of $G_\maximevar$. Unlike \rubenvar, \maximevar\ never turns a reifying triple into an edge of its own; each reifying triple instead becomes a single edge recording its reifier and its assertion status. Here this creates six nodes: $\ntfun(\isun)$, $\ntfun(\ipopcorn)$, $\ntfun(\iplasma)$, and $\ntfun(\ialice)$ labeled $\mathsf{IRI}$, and $\ntfun(r_1)$, $\ntfun(r_2)$ labeled $\mathsf{Reifier}$; and three edges: the edge for $\tau_1$ with $\kasserted\mapsto\mathtt{false}$; the edge for $\tau_2$, created for $r_2$ and then flipped to $\kasserted\mapsto\mathtt{true}$ by the plain triple $(\isun,\imadeof,\iplasma)$; and the asserted edge $(r_1,\istatedby,\ialice)$ for the description. Each of the first two edges references its reifier node, $r_1$ and $r_2$ respectively. Figure~\ref{fig:maxime-example} shows the output $\maximevar(G_\maximevar)$.
\end{example}

When a triple term edge is created, it is not yet known whether it is asserted, since the triple $\tau$ may also occur as a plain triple of $\graph$. Algorithm~\ref{alg:maxime} therefore proceeds in two passes. Pass~1 creates, for each reifying triple, one triple term edge carrying $\kasserted \mapsto \mathtt{false}$, and records it in the map $\textit{ttEdges}$.\footnote{Note that a triple term can have multiple reifiers. Here, this results in parallel triple term edges (one per reifying triple).} Pass~2 considers each top-level triple $(s,p,o)$. If $\textit{ttEdges}$ already holds edges for $(s,p,o)$, then that triple is asserted after all and every such edge is flipped to $\kasserted \mapsto \mathtt{true}$; otherwise $(s,p,o)$ is an ordinary triple and becomes a new edge carrying $\kasserted \mapsto \mathtt{true}$. RDF~1.2 treats the occurrence of a triple as a triple term or top-level triple independently. Representing both occurrences using a single LPG edge carrying the $\kasserted$ property is therefore a design choice of the translation.
 
\begin{algorithm}[h]
  \small
\caption{\maximevar: reifies-interpreting translation}
\label{alg:maxime}
\begin{algorithmic}[1]
  \Statex \textbf{Shared state:} $N, E, \edgf, \labf, \recf$
  \Procedure{\maximevar}{$\graph$}
    \State $N \gets \emptyset$;\quad $E \gets \emptyset$ ;\quad
    $\edgf, \labf, \recf \gets$ empty maps
    \State $\textit{ttEdges} \gets$ empty map
    \Comment{maps triples to triple term edges}
    \Statex \emph{Pass 1: each reifying triple creates one triple term edge}
    \ForAll{$(r, \ireifies, \tau) \in \graph$ \textbf{with} $\tau \in \tterms$}
        \State $\textsc{AddReifier}(r)$
      \State $e \gets \textsc{CreateEdge}\big(\tau,\
      \{\kasserted\mapsto \mathtt{false},\
      \kreifiedby\mapsto \ntfun(r)\}\big)$
      \State $\textit{ttEdges}(\tau) \gets \textit{ttEdges}(\tau) \cup \{e\}$
    \EndFor
    \Statex \emph{Pass 2: plain triples assert triple term edges or create new edges}
    \ForAll{$(s, p, o) \in \graph$ \textbf{with} $o \notin \tterms$}
      \If{$\textit{ttEdges}(s,p,o) \neq \emptyset$}
        \Comment{$(s,p,o)$ is a triple term edge}
        \ForAll{$e \in \textit{ttEdges}(s,p,o)$}
          \State $\recf(e)(\kasserted) \gets \mathtt{true}$
          \Comment{triple term edge asserted iff its triple is in $\graph$}
        \EndFor
      \Else
        \State \textsc{CreateEdge}$\big((s,p,o),\ \{\kasserted\mapsto \mathtt{true}\}\big)$
      \EndIf
    \EndFor
    \State \Return $(N, E, \edgf, \labf, \recf)$
  \EndProcedure
 
  \Statex
  \Procedure{CreateEdge}{$(s, p, o),\ \textit{rec}$}
      \Comment{one edge carrying record \textit{rec}}
    \State \textsc{AddNode}$(s)$;\quad \textsc{AddNode}$(o)$
      \Comment{\textsc{AddNode} as in Algorithm~\ref{alg:ruben}}
    \State choose $e \in \edges \setminus E$;\quad $E \gets E \cup \{e\}$
    \State $\edgf(e) \gets \big(\ntfun(s),\ \ntfun(o)\big)$;\quad
           $\labf(e) \gets \{\ntfun(p)\}$
    \State $\recf(e) \gets \textit{rec}$
    \State \Return $e$
  \EndProcedure
 
  \Statex
  \Procedure{AddReifier}{$r$}
      \Comment{materialize a reifier as a node}
    \If{$\ntfun(r) \notin N$}
      \State $N \gets N \cup \{\ntfun(r)\}$
      \State $\labf(\ntfun(r)) \gets \{\mathsf{Reifier}\}$
      \State $\recf(\ntfun(r)) \gets \textsc{Record}(r)$
        \Comment{carries $\krefid\mapsto\ntfun(r)$}
    \EndIf
  \EndProcedure
\end{algorithmic}
\end{algorithm}

\subsection{Translation \katjavar} \label{sec:katjavar}
\katjavar\ considers flat reification graphs (cf.~Section~\ref{sec:maximevar}) satisfying two further assumptions: reifiers occur in subject position only, and each reifier reifies a single triple term (several reifiers per triple term remain allowed); we call such graphs \emph{strict} flat (reification) graphs. Without the latter assumption, Pass~1 below would fold a reifier's description onto every triple term it reifies. Like \maximevar, \katjavar\ interprets $\ireifies$ and represents a reified triple $\tau=(s,p,o)$ as a triple term edge from $s$ to $o$. It differs in what it does with the reifier. While \maximevar\ turns a reifier into a node, \katjavar\ never makes a reifier a node. Instead, descriptions for $\tau$ (i.e., triples where the subject is a reifier cf. Section~\ref{sec:prelim}) are stored in the record of the triple term edges. We call the process of storing descriptions in edge records ``folding.'' A reifier therefore is not a node of its own, and descriptions are folded into the triple term edge for $\tau$.

Two folding strategies are offered. \emph{Fixed folding} uses two reserved keys: $\krefprop$ for the description predicate, and $\kreferences$ for a foreign key to the description object. As these keys are fixed, one edge holds a single description, so a reifier with several descriptions requires one edge per description. \emph{Open folding} instead uses each description predicate itself as a property key, with the object's foreign key as its record value, so that all descriptions of a reifier fit in a single triple term edge. This is more concise, but requires two restrictions: predicate identifiers must serve as property keys ($\str\subseteq\keys$), so the record has no predetermined key set; and since a record maps each key to one value, no two descriptions of the same reifier may share a predicate.\footnote{Repeated predicates could be handled by  list-valued properties. To keep our algorithms concise, we do not formalize or implement this detail.} In both strategies the description object is represented as a node, addressable through the foreign key in the triple term edge.

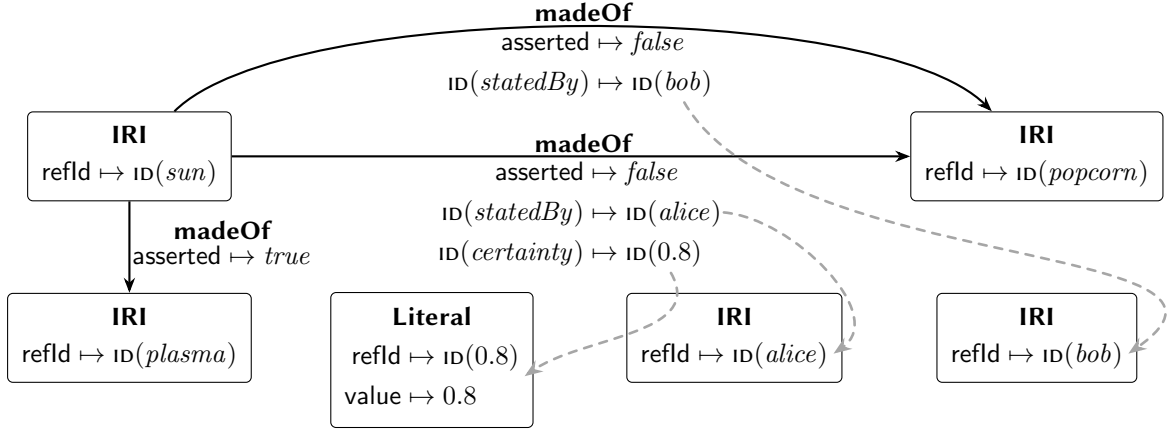
\begin{figure}
\centering
\begin{tikzpicture}[x=1cm, y=0.8cm]
 
  \gnode{plasma} {0,0}{IRI}     {$\krefids\mapsto \ntfun(\iplasma)$}
  \gnode{sun}    {0,3}{IRI}     {$\krefids\mapsto \ntfun(\isun)$}
  \gnode{popcorn}{12,3}{IRI}    {$\krefids\mapsto \ntfun(\ipopcorn)$}
  \gnode{alice}{8,0}{IRI}    {$\krefids\mapsto \ntfun(\ialice)$}
  \gnode{bob}{12,0}{IRI}    {$\krefids\mapsto \ntfun(\ibob)$}
  \gnode{value}{4,0}{Literal}    {\recl{
      \krefids &\mapsto \ntfun(0.8)\\
      \mathsf{value} &\mapsto 0.8\\}}
  
  \draw[gedge] (sun) -- (popcorn);
  \draw[gedge] (sun) to[out=45, in=135, looseness=0.5]
  coordinate[pos=0.5] (e2mid) (popcorn);
  \draw[gedge] (sun) -- (plasma);
  
  \gedgelab      {e1}{6,1}{madeOf}{
    \recl{
      \kasserted &\mapsto \mathit{false}\\
      \ntfun(\istatedby) &\mapsto\ntfun(\ialice)\\
      \ntfun(\icertainty) &\mapsto\ntfun(0.8)\\
    }
    }
    \gedgelab      {e2}{[yshift=-1cm]e2mid}{madeOf}{
      \recl{
        \kasserted &\mapsto \mathit{false}\\
        \ntfun(\istatedby) &\mapsto\ntfun(\ibob)\\
      }
    }
    \gedgelab[west]      {e3}{0,1.3}{madeOf}{$\kasserted\mapsto \mathit{true}$}
    
  \begin{pgfonlayer}{annot}
    \draw[ghost,dashed] ([xshift=-1mm,yshift=8mm] e1rec.south east) to[out=-5, in=35] (alicerec.east);
    \draw[ghost,dashed] ([xshift=-5mm] e2rec.south east) to[out=-60, in=35] (bobrec.east);
    \draw[ghost,dashed] ([xshift=-8mm] e1rec.south east) to[out=-60, in=35] (valuerec.east);
  \end{pgfonlayer}
\end{tikzpicture}
\caption{The \katjavar\ encoding of Example~\ref{ex:katja} with open folding.}
\label{fig:katja-example}
\end{figure}

\begin{example}\label{ex:katja}
  We extend $G_\maximevar$ of Example~\ref{ex:maxime} with the second   description of $r_1$ and a second reifier $r_3$ of $\tau_1$, completing  the graph of Figure~\ref{lst:intro}:
  $G_\katjavar = G_\maximevar \cup
  \{(r_1,\icertainty,0.8),\,(r_3,\ireifies,\tau_1),\,(r_3,\istatedby,\ibob)\}$.

  Figure~\ref{fig:katja-example} shows $\katjavar(G_\katjavar)$ under open folding. None of the three description triples becomes an edge, and no reifier becomes a node: their content is folded in Pass~1, and Pass~2 skips triples mentioning reifiers---the essential difference from \maximevar, where $(r_1,\istatedby,\ialice)$ was an edge of its own. The six nodes are the five IRIs and a $\mathsf{Literal}$ node for $0.8$: every description object is materialized so that the foreign keys in the folded records can reach it.
    
  Under open folding, each reifying triple contributes one triple term   edge. The two reifiers of $\tau_1$ thus yield two parallel $\imadeof$  edges from $\ntfun(\isun)$ to $\ntfun(\ipopcorn)$, distinguished only  by their folded records---which retain the descriptions ($\ialice$,  $\ibob$, $0.8$) but no trace of $r_1$ or $r_3$ themselves. Both keep  $\kasserted\mapsto\mathtt{false}$, as $\tau_1$ is never asserted; the  edge for $\tau_2$, created for the description-free $r_2$, is flipped  to $\mathtt{true}$ by Pass~2 as in Example~\ref{ex:maxime}. Under fixed folding, the two descriptions of $r_1$ would instead occupy one edge each (via $\krefprop$/$\kreferences$), giving four triple term edges rather than three.
\end{example}

Algorithm~\ref{alg:katja} also defines two passes. Pass~1 folds the descriptions of each reifier onto triple term edges for $\tau$, all initially carrying $\kasserted\mapsto\mathtt{false}$: fixed folding creates one such edge per description, whereas open folding creates a single edge carrying every description. A reifier with no description creates one triple term edge carrying only $\kasserted\mapsto\mathtt{false}$. Pass~2 then visits every top-level triple whose subject and object are not reifiers. If it corresponds to a triple term edge, assertedness is flipped to $\kasserted\mapsto\mathtt{true}$; otherwise it becomes a regular (asserted) edge. A triple that mentions a reifier is skipped. Because a reifier is never a node, its descriptions were already folded in Pass~1.

\begin{algorithm}[h]
  \small
\caption{\katjavar: reifies-interpreting translation with folded descriptions}
\label{alg:katja}
\begin{algorithmic}[1]
  \Statex \textbf{Shared state:} $N, E, \edgf, \labf, \recf$;\quad
          \textbf{parameter} $\mathit{form} \in \{\mathit{fixed}, \mathit{open}\}$
  \Procedure{\katjavar}{$\graph$}
    \State $N \gets \emptyset$;\quad $E \gets \emptyset$;\quad
           $\edgf, \labf, \recf \gets$ empty maps
    \State $\textit{reifiers} \gets
           \{\, r \mid (r, \ireifies, \tau) \in \graph,\ \tau \in \tterms \,\}$
    \State $\textit{ttEdges} \gets$ empty map
      \Comment{maps triples to triple term edges}
    \Statex \emph{Pass 1: fold each reifier's descriptions onto triple term edges}
    \ForAll{$(r, \ireifies, \tau) \in \graph$ \textbf{with} $\tau \in \tterms$}
      \State $\textit{descriptions} \gets
             \{\, (r, q, v) \in \graph \mid q \neq \ireifies \,\}$
      \ForAll{$(r, q, v) \in \textit{descriptions}$}
        \State \textsc{AddNode}$(v)$
          \Comment{description object becomes a node}
      \EndFor
      \ForAll{$\textit{rec} \in \textsc{CreateRecords}(\textit{descriptions})$}
        \State $e \gets \textsc{CreateEdge}(\tau,\ \textit{rec})$
        \State $\textit{ttEdges}(\tau) \gets \textit{ttEdges}(\tau) \cup \{e\}$
      \EndFor
    \EndFor
    \Statex \emph{Pass 2: plain triples that mention no reifier}
    \ForAll{$(s, p, o) \in \graph$ \textbf{with}
            $o \notin \tterms$ \textbf{and} $s, o \notin \textit{reifiers}$}
      \If{$\textit{ttEdges}(s,p,o) \neq \emptyset$}
        \ForAll{$e \in \textit{ttEdges}(s,p,o)$}
          \State $\recf(e)(\kasserted) \gets \mathtt{true}$
        \EndFor
      \Else
        \State \textsc{CreateEdge}$((s,p,o),\ \{\kasserted\mapsto\mathtt{true}\})$
      \EndIf
    \EndFor
    \State \Return $(N, E, \edgf, \labf, \recf)$
  \EndProcedure

  \Statex
  \Procedure{CreateRecords}{$\textit{descriptions}$}
      \Comment{set of edge records for a reifier's descriptions, per \textit{form}}
    \If{$\mathit{form} = \mathit{open}$}
        \Comment{all descriptions on one record}
      \State $\textit{rec} \gets \{\kasserted\mapsto\mathtt{false}\}$
      \ForAll{$(r, q, v) \in \textit{descriptions}$}
        \State $\textit{rec}(\ntfun(q)) \gets \ntfun(v)$
          \Comment{predicate is the key}
      \EndFor
      \State \Return $\{\, \textit{rec} \,\}$
    \EndIf
    \If{$\textit{descriptions} = \emptyset$}
      \State \Return $\{\ \{\kasserted\mapsto\mathtt{false}\}\ \}$
        \Comment{fixed, bare reifier}
    \EndIf
    \State \Return $\{\ \{\kasserted\mapsto\mathtt{false},\
           \krefprop\mapsto \ntfun(q),\
           \kreferences\mapsto \ntfun(v)\}
           \mid (r,q,v) \in \textit{descriptions}\,\}$
        \Comment{fixed, one record per description}
  \EndProcedure
\end{algorithmic}
\end{algorithm}

\subsection{Trade-offs}
\label{sec:tradeoffs}

Table~\ref{tab:tradeoffs} summarizes the trade-off between completeness (i.e., which graphs a translation accepts, and what it preserves) and how closely its output realizes LPGs' native edge-with-properties construct. Call a translation \emph{lossless} on a class of RDF graphs if the input can be reconstructed exactly from the output LPG; as node identifiers can be read back to RDF terms, this hinges only on the four rows of Table~\ref{tab:tradeoffs}, that is, how triple terms, reifiers, descriptions, and assertedness are encoded.
\rubenvar\ is lossless on \emph{every} RDF~1.2 graph: an edge without $\kin$ is an asserted triple, and a $\mathsf{TripleTerm}$ node's triple is recovered, recursively, through its $\kin$ edge.
\maximevar\ is lossless on flat reification graphs, the class delimited by its two assumptions: each edge carrying $\kreifiedby$ decodes to its reifying triple, plus the plain triple iff $\kasserted$ is $\mathtt{true}$, and every other edge (descriptions included) decodes directly to its triple.
\katjavar\ is lossy even on strict flat reification graphs, precisely at the reifiers: renaming $r_1$ in Figure~\ref{lst:intro} gives us the same LPG (the output in Example~\ref{ex:katja} carries no trace of $r_1$ or $r_3$), and a bare reifier of an asserted triple leaves no trace at all. Beyond these two effects nothing is lost under open folding; fixed folding additionally forgets which descriptions share a reifier, in exchange for its fixed record schema, where open folding's keys are data-dependent (Section~\ref{sec:katjavar}). Size in Table~\ref{tab:tradeoffs} counts the elements (nodes and edges) of the output on inputs all variants accept.

\begin{table}[t]
  \small
  \caption{Overview. \emph{Flat graphs}: triple terms occur only as objects of $\ireifies$ and are not nested; \emph{strict}: additionally, reifiers occur only as subjects, each reifying a single triple term.}
  \label{tab:tradeoffs}
  \begin{tabular}{@{}lp{3cm}p{3.2cm}p{3.6cm}@{}}
    \toprule
    & \rubenvar & \maximevar & \katjavar \\
    \midrule
    Input graphs & all of RDF~1.2 & flat graphs & strict flat graphs \\
    Triple term & node + content edge & edge per reifying triple & edge(s) per reifying triple \\
    Reifier & ordinary node & $\mathsf{Reifier}$ node & absent \\
    Descriptions & ordinary edges & ordinary edges & folded into edge records \\
    Assertedness & absence of $\kin$ & $\kasserted$ key & $\kasserted$ key \\
    Lossless & yes & yes, on its class & no: reifiers lost \\
    Size (nodes + edges) & largest & intermediate & smallest \\
    \bottomrule
  \end{tabular}
\end{table}

\section{Experimental Evaluation}
\label{sec:evaluation}
We implemented all three translation algorithms of Section~\ref{sec:translations}; for \katjavar, fixed and open folding are measured separately. The evaluation has two goals: (1)~to demonstrate empirically that each variant converts RDF~1.2 data into labeled property graphs on its intended class of inputs, and (2)~to test how the trade-offs predicted in Section~\ref{sec:tradeoffs} materialize in practice. We measure conversion time, the converter's memory usage (peak Resident Set Size (RSS)), and the size of the generated output.
 
For brevity, we omit methodological details (hardware and software configuration, repetitions per configuration, etc.); all reported numbers are averages over repeated runs. An extended version of this section, the raw results of every measurement, and reproduction instructions are available in the accompanying repository\footnote{\url{https://github.com/RDF-PG-Interoperability/rdf12-to-pg/tree/cc328d1733502dabbe964f63551795ad9436b7d6/experiments}}.
 
\paragraph{Datasets}
Our main workload consists of samples of the RDF-star representation of the Biomedical Knowledge Repository (BKR) published with the REF benchmark
\cite{orlandi_2020_4148888} 
and used by StarBench
\cite{abouda2023starbench}
; we evaluate conversion only, not query execution. In BKR, each embedded triple is a biomedical proposition annotated with \emph{provenance} values: the sources from which the proposition derives. As the source uses the earlier RDF-star syntax, in which an embedded triple may occupy subject position, each record \verb|<<s p o>> q v1, ..., vk| is rewritten with a fresh, deterministically named blank-node reifier $r$ as \verb|r rdf:reifies <<(s p o)>>| and \verb|r q v1, ..., vk|. This retains the embedded triple's terms, its non-asserted status, and its provenance values, but changes the graph structure and query patterns.
 
We sampled five target sizes, from 0.5\,MB to 2.5\,MB in 0.5\,MB increments (Table~\ref{tab:bkr-datasets}). The samples are deterministic but not nested prefixes of one another, so their composition varies with size. After migration, every provenance value becomes a description of its proposition's reifier, so the samples exhibit a high description count of 57 to 177 descriptions per reifier; we call such workloads \emph{description-heavy}. 
 
\begin{table}[t]
\centering
\small
\caption{Samples derived from the BKR/REF RDF-star data and migrated to RDF~1.2.}
\label{tab:bkr-datasets}
\begin{tabular}{lrrrrr}
\toprule
Target size & Reifiers & Descriptions & Descr./reifier & Turtle bytes & Graph triples\\
\midrule
0.5~MB & 154 & 8,835 & 57.4 & 499,932 & 8,989\\
1~MB & 134 & 18,554 & 138.5 & 999,917 & 18,688\\
1.5~MB & 159 & 28,042 & 176.4 & 1,499,694 & 28,201\\
2~MB & 211 & 37,402 & 177.3 & 1,999,851 & 37,613\\
2.5~MB & 286 & 46,639 & 163.1 & 2,499,756 & 46,925\\
\bottomrule
\end{tabular}
\end{table}
 
We additionally ran the implementations on a synthetic dataset covering many RDF~1.2 features, generated by a publicly available tool\footnote{\url{https://github.com/RDF-PG-Interoperability/rdf12-to-pg/tree/cc328d1733502dabbe964f63551795ad9436b7d6/synthetic-generator}}. It includes standalone and nested triple terms, which fall outside the input restrictions of \maximevar\ and \katjavar\ (Section~\ref{sec:translations}); their implementations skip such triples unless the predicate is \texttt{rdf:reifies}. Successful conversion of this workload therefore demonstrates operational robustness, not information preservation outside the stated domains. Details on these additional experiments can be found in the repository.
 
\paragraph{Output formats}
Each conversion outputs YARS-PG syntax~\cite{tomaszuk2019serialization} or a Neo4j loading script in Cypher, either \emph{monolithic} (the entire graph in one transaction) or \emph{batched} (1,000 rows per transaction, with a loader-specific indexed identifier that is not part of the conceptual translations). Every combination of translation, folding form, output format, and input size was executed.

\begin{figure}[t]
\centering
\begingroup

\newcommand{\bkrBaseBar}[3]{%
  \addplot+[forget plot,draw=black,fill=#1,bar shift=#2] coordinates {#3};%
}
\newcommand{\bkrSolidBar}[3]{\bkrBaseBar{#1}{#2}{#3}}
\newcommand{\bkrDottedBar}[3]{%
  \bkrBaseBar{#1}{#2}{#3}%
  \addplot+[forget plot,draw=none,fill=none,pattern=dots,
    pattern color=black,bar shift=#2] coordinates {#3};%
}
\newcommand{\bkrHatchedBar}[3]{%
  \bkrBaseBar{#1}{#2}{#3}%
  \addplot+[forget plot,draw=none,fill=none,pattern=north east lines,
    pattern color=black,bar shift=#2] coordinates {#3};%
}

\begin{tikzpicture}
\begin{axis}[
  ybar,
  bar width=4.5pt,
  width=\textwidth,
  height=4.5cm,
  ymin=0,
  ymax=5,
  ylabel={Time (s)},
  symbolic x coords={0.5MB,1MB,1.5MB,2MB,2.5MB},
  xtick=data,
  enlarge x limits=0.10,
  scaled y ticks=false
]

\bkrSolidBar{green}{-27.5pt}{
  (0.5MB,0.230) (1MB,0.441) (1.5MB,0.646) (2MB,0.876) (2.5MB,1.103)
}
\bkrDottedBar{green}{-22.5pt}{
  (0.5MB,0.232) (1MB,0.441) (1.5MB,0.658) (2MB,0.885) (2.5MB,1.113)
}
\bkrHatchedBar{green}{-17.5pt}{
  (0.5MB,0.662) (1MB,1.421) (1.5MB,2.309) (2MB,3.311) (2.5MB,4.453)
}
\bkrSolidBar{red}{-12.5pt}{
  (0.5MB,0.231) (1MB,0.452) (1.5MB,0.664) (2MB,0.899) (2.5MB,1.143)
}
\bkrDottedBar{red}{-7.5pt}{
  (0.5MB,0.240) (1MB,0.461) (1.5MB,0.682) (2MB,0.917) (2.5MB,1.152)
}
\bkrHatchedBar{red}{-2.5pt}{
  (0.5MB,0.281) (1MB,0.537) (1.5MB,0.792) (2MB,1.050) (2.5MB,1.318)
}
\bkrSolidBar{orange}{2.5pt}{
  (0.5MB,0.219) (1MB,0.413) (1.5MB,0.608) (2MB,0.803) (2.5MB,1.011)
}
\bkrDottedBar{orange}{7.5pt}{
  (0.5MB,0.220) (1MB,0.410) (1.5MB,0.617) (2MB,0.813) (2.5MB,1.021)
}
\bkrHatchedBar{orange}{12.5pt}{
  (0.5MB,0.261) (1MB,0.496) (1.5MB,0.746) (2MB,0.997) (2.5MB,1.241)
}
\bkrSolidBar{blue}{17.5pt}{
  (0.5MB,0.220) (1MB,0.421) (1.5MB,0.613) (2MB,0.824) (2.5MB,1.043)
}
\bkrDottedBar{blue}{22.5pt}{
  (0.5MB,0.223) (1MB,0.423) (1.5MB,0.622) (2MB,0.848) (2.5MB,1.060)
}
\bkrHatchedBar{blue}{27.5pt}{
  (0.5MB,0.488) (1MB,1.080) (1.5MB,1.779) (2MB,2.568) (2.5MB,3.523)
}
\end{axis}
\end{tikzpicture}

\begin{tikzpicture}
\begin{axis}[
  ybar,
  bar width=4.5pt,
  width=\textwidth,
  height=4.5cm,
  ymin=0,
  ymax=150,
  ylabel={Peak RSS (MiB)},
  symbolic x coords={0.5MB,1MB,1.5MB,2MB,2.5MB},
  xtick=data,
  enlarge x limits=0.10,
  scaled y ticks=false
]

\bkrSolidBar{green}{-27.5pt}{
  (0.5MB,35.502) (1MB,57.341) (1.5MB,81.971) (2MB,100.943) (2.5MB,127.880)
}
\bkrDottedBar{green}{-22.5pt}{
  (0.5MB,40.029) (1MB,62.230) (1.5MB,87.368) (2MB,106.463) (2.5MB,134.010)
}
\bkrHatchedBar{green}{-17.5pt}{
  (0.5MB,25.005) (1MB,30.039) (1.5MB,37.120) (2MB,40.271) (2.5MB,50.017)
}
\bkrSolidBar{red}{-12.5pt}{
  (0.5MB,36.885) (1MB,60.792) (1.5MB,85.690) (2MB,107.077) (2.5MB,135.731)
}
\bkrDottedBar{red}{-7.5pt}{
  (0.5MB,41.278) (1MB,65.586) (1.5MB,91.696) (2MB,112.822) (2.5MB,141.141)
}
\bkrHatchedBar{red}{-2.5pt}{
  (0.5MB,28.630) (1MB,39.703) (1.5MB,50.491) (2MB,56.520) (2.5MB,68.592)
}
\bkrSolidBar{orange}{2.5pt}{
  (0.5MB,29.399) (1MB,45.660) (1.5MB,62.597) (2MB,75.314) (2.5MB,95.557)
}
\bkrDottedBar{orange}{7.5pt}{
  (0.5MB,33.852) (1MB,50.099) (1.5MB,67.343) (2MB,80.209) (2.5MB,100.665)
}
\bkrHatchedBar{orange}{12.5pt}{
  (0.5MB,33.246) (1MB,48.940) (1.5MB,66.050) (2MB,78.925) (2.5MB,99.307)
}
\bkrSolidBar{blue}{17.5pt}{
  (0.5MB,33.424) (1MB,53.228) (1.5MB,74.893) (2MB,92.185) (2.5MB,115.371)
}
\bkrDottedBar{blue}{22.5pt}{
  (0.5MB,38.054) (1MB,58.175) (1.5MB,80.035) (2MB,97.883) (2.5MB,123.362)
}
\bkrHatchedBar{blue}{27.5pt}{
  (0.5MB,24.784) (1MB,29.789) (1.5MB,36.641) (2MB,39.699) (2.5MB,49.081)
}
\end{axis}
\end{tikzpicture}

\begin{tikzpicture}
\begin{axis}[
  ybar,
  bar width=4.5pt,
  width=\textwidth,
  height=4.5cm,
  ymin=0,
  ymax=12,
  ylabel={Output size (MB)},
  symbolic x coords={0.5MB,1MB,1.5MB,2MB,2.5MB},
  xtick=data,
  enlarge x limits=0.10,
  scaled y ticks=false
]

\bkrSolidBar{green}{-27.5pt}{
  (0.5MB,1.444) (1MB,3.027) (1.5MB,4.570) (2MB,6.039) (2.5MB,7.550)
}
\bkrDottedBar{green}{-22.5pt}{
  (0.5MB,1.533) (1MB,3.213) (1.5MB,4.850) (2MB,6.409) (2.5MB,8.012)
}
\bkrHatchedBar{green}{-17.5pt}{
  (0.5MB,1.582) (1MB,3.287) (1.5MB,4.969) (2MB,6.631) (2.5MB,8.285)
}
\bkrSolidBar{red}{-12.5pt}{
  (0.5MB,2.067) (1MB,4.397) (1.5MB,6.652) (2MB,8.811) (2.5MB,10.997)
}
\bkrDottedBar{red}{-7.5pt}{
  (0.5MB,2.154) (1MB,4.581) (1.5MB,6.930) (2MB,9.178) (2.5MB,11.455)
}
\bkrHatchedBar{red}{-2.5pt}{
  (0.5MB,2.173) (1MB,4.577) (1.5MB,6.929) (2MB,9.243) (2.5MB,11.531)
}
\bkrSolidBar{orange}{2.5pt}{
  (0.5MB,0.620) (1MB,1.295) (1.5MB,1.945) (2MB,2.530) (2.5MB,3.167)
}
\bkrDottedBar{orange}{7.5pt}{
  (0.5MB,0.655) (1MB,1.369) (1.5MB,2.056) (2MB,2.675) (2.5MB,3.348)
}
\bkrHatchedBar{orange}{12.5pt}{
  (0.5MB,0.710) (1MB,1.483) (1.5MB,2.227) (2MB,2.895) (2.5MB,3.624)
}
\bkrSolidBar{blue}{17.5pt}{
  (0.5MB,1.277) (1MB,2.663) (1.5MB,4.019) (2MB,5.303) (2.5MB,6.635)
}
\bkrDottedBar{blue}{22.5pt}{
  (0.5MB,1.376) (1MB,2.869) (1.5MB,4.329) (2MB,5.713) (2.5MB,7.147)
}
\bkrHatchedBar{blue}{27.5pt}{
  (0.5MB,1.453) (1MB,3.006) (1.5MB,4.540) (2MB,6.059) (2.5MB,7.569)
}
\end{axis}

\draw[draw=black,fill=green] (0.20,-1.05) rectangle +(0.32,0.22);
\node[anchor=west,font=\small] at (0.67,-0.94) {\maximevar};
\draw[draw=black,fill=red] (3.25,-1.05) rectangle +(0.32,0.22);
\node[anchor=west,font=\small] at (3.72,-0.94) {\katjavar\ fixed};
\draw[draw=black,fill=orange] (6.65,-1.05) rectangle +(0.32,0.22);
\node[anchor=west,font=\small] at (7.12,-0.94) {\katjavar\ open};
\draw[draw=black,fill=blue] (10.10,-1.05) rectangle +(0.32,0.22);
\node[anchor=west,font=\small] at (10.57,-0.94) {\rubenvar};

\draw[draw=black] (1.00,-1.70) rectangle +(0.32,0.22);
\node[anchor=west,font=\small] at (1.47,-1.59) {YARS-PG};
\draw[draw=black] (5.00,-1.70) rectangle +(0.32,0.22);
\fill[pattern=dots,pattern color=black] (5.00,-1.70) rectangle +(0.32,0.22);
\node[anchor=west,font=\small] at (5.47,-1.59) {Cypher monolithic};
\draw[draw=black] (9.70,-1.70) rectangle +(0.32,0.22);
\fill[pattern=north east lines,pattern color=black] (9.70,-1.70) rectangle +(0.32,0.22);
\node[anchor=west,font=\small] at (10.17,-1.59) {Cypher batched};
\end{tikzpicture}

\endgroup
\caption{Conversion time, peak RSS, and output size for the BKR dataset; fixed and open folding are shown separately.}
\label{fig:bkr_experiments}
\end{figure}
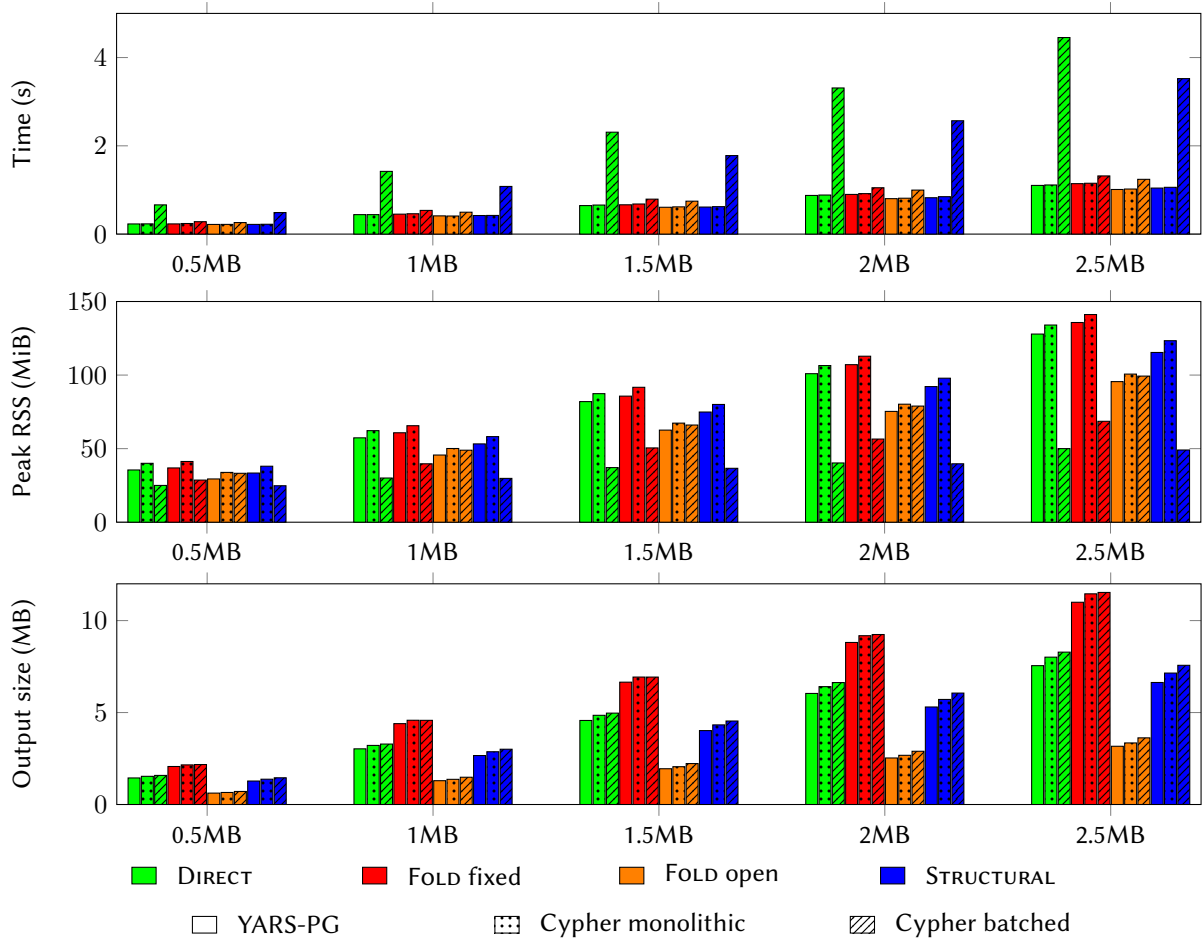

\paragraph{Results}
All runs completed without errors. Figure~\ref{fig:bkr_experiments} shows conversion time, peak RSS, and output size on the BKR workload. Four observations stand out.

First, all measurements grow approximately linearly, and monolithic conversion time is nearly identical across the algorithms (1.02--1.15\,s at 2.5\,MB): the generality of \rubenvar\ has no discernible runtime penalty.

Second, the choice of output format matters more than the choice of translation. The batched scripts are designed for memory-constrained loading, and generating is a speed vs. memory trade-off: the converter tracks already-created nodes on disk rather than in memory, which for \maximevar\ and \rubenvar\ slows generation (4.5\,s and 3.5\,s at 2.5\,MB, versus roughly one second in monolithic mode) and enlarges the files, but more than halves their memory usage ($\approx$50\,MiB versus 124--134\,MiB). The \katjavar\ implementation does less of this bookkeeping, so its batched generation stays fast (1.2--1.3\,s), with correspondingly smaller memory savings.

Third, the folding strategy dominates output size on this description-heavy workload. \maximevar, \rubenvar, and fixed folding all produce one edge per description, whereas open folding produces one triple term edge per reifier (at 2.5\,MB, 286 edges in place of 46,639). Byte sizes (Table~\ref{tab:bkr-output-size}), however, do not simply follow these counts, since what each element records matters as much. Fixed folding, though smaller in nodes and edges than \maximevar\ and \rubenvar, produces the largest files: every per-description edge repeats the reified triple's source, target, and label, which at 57--177 descriptions per reifier outweighs the reifier nodes and description edges it saves. The same effect ranks \rubenvar\ below \maximevar: \maximevar\ marks every edge with the $\kasserted$ flag, which on tens of thousands of edges outweighs the few hundred extra nodes \rubenvar\ creates. Open folding results by far in the smallest output, though its measured bytes should be interpreted with caution: BKR's descriptions share one provenance predicate per reifier, and our prototype retains a single value per predicate key.\footnote{Instead, one could implement the list-valued extension of Section~\ref{sec:katjavar} to fully preserve information.}

\begin{table}[t]
\centering
\small
\caption{Output size at the 2.5\,MB input (monolithic loading, in MB).}
\label{tab:bkr-output-size}
\begin{tabular}{lrrrr}
\toprule
 & \maximevar & \katjavar\ fixed & \katjavar\ open & \rubenvar\\
\midrule
Cypher  & 8.01 & 11.46 & 3.35 & 7.15\\
YARS-PG & 7.55 & 11.00 & 3.17 & 6.63\\
\bottomrule
\end{tabular}
\end{table}

Fourth, complementary measurements on the synthetic workload, reported in the repository, suggest that the batched scripts pay off downstream: in the one setting measured in both modes, loading the batched script into Neo4j took 6.97\,s on average, compared with 345.16\,s for the monolithic script, making it roughly 50 times faster.

In total, the experiments demonstrate the feasibility of all three translations and support the trade-offs discussed in Section~\ref{sec:tradeoffs}.
 
\section{Conclusion}
\label{sec:conclusion}
We studied the translation of RDF~1.2 graphs into LPGs, focusing on the statement-level constructs that LPGs lack. Our three translations differ in how much of the reification vocabulary they interpret: \rubenvar\ interprets none of it and is lossless on all of RDF~1.2; \maximevar\ interprets \texttt{rdf:reifies} and is lossless on flat reification graphs while retaining reifiers as nodes;
and \katjavar\ folds descriptions into edge records, the native LPG representation of annotated statements, at the cost of reifier identity.
Our experiments show that all three are practical and corroborate the predicted trade-offs: the generality of \rubenvar\ carries no discernible runtime penalty, making it the safest general-purpose choice; open folding makes \katjavar\ the variant of choice for description-heavy data, producing one annotated edge per reifier where the other configurations produces one per description.

Several directions remain open. A natural next step is to complement structural losslessness with \emph{query losslessness}: translating SPARQL~1.2 queries into GQL queries over the translated LPG and validating answer equivalence, thereby characterizing each translation by the queries it preserves rather than the graphs it reconstructs. On the representation side, literal-valued descriptions could be inlined as plain record values rather than materialized as referenceable nodes, improving both conciseness and performance. Finally, we plan to lift the translations from RDF graphs to RDF datasets with named-graph support, and to revisit aspects
left out of scope here: entailment, native blank-node handling instead of Skolemization, and vendor-specific features of concrete graph database systems.
 

\section*{Acknowledgments}
This publication is based upon work performed during a hackathon of COST Action CA23147 GOBLIN --- Global Network on Large-Scale, Cross-domain and Multilingual Open Knowledge Graphs, supported by COST (European Cooperation in Science and
Technology).

\section*{Declaration on Generative AI}
During the preparation of this work, the authors used Claude Fable 5 for: Drafting content, Peer review simulation, Content enhancement, and Formatting assistance. The authors extensively reviewed and edited the content as needed and assume full responsibility for the content of the publication.

\bibliography{references}


\end{document}